\documentclass[prl,twocolumn,showpacs,aps,superscriptaddress,floatfix]{revtex4}
\usepackage{amsmath}
\usepackage{amssymb}
\usepackage{bm}
\usepackage{graphicx}
\usepackage{color}
\usepackage{hyperref}
\usepackage{verbatim}

\begin{document}

\title{Spin-valley half-metal as a prospective material for spin-valley-tronics}

\author{A.V. Rozhkov}
\affiliation{Center for Emergent Matter Science, RIKEN, Wako-shi, Saitama,
351-0198, Japan}
\affiliation{Institute for Theoretical and Applied Electrodynamics, Russian Academy of Sciences, Moscow, 125412 Russia}
\affiliation{Moscow Institute for Physics and Technology (State
University), Dolgoprudnyi, 141700 Russia}

\author{A.L. Rakhmanov}
\affiliation{Center for Emergent Matter Science, RIKEN, Wako-shi, Saitama,
351-0198, Japan}
\affiliation{Moscow Institute for Physics and Technology (State
University), Dolgoprudnyi, 141700 Russia}
\affiliation{Institute for Theoretical and Applied Electrodynamics, Russian Academy of Sciences, Moscow, 125412 Russia}
\affiliation{Dukhov Research Institute of Automatics, Moscow, 127055 Russia}

\author{A.O. Sboychakov}
\affiliation{Center for Emergent Matter Science, RIKEN, Wako-shi, Saitama,
351-0198, Japan}
\affiliation{Institute for Theoretical and Applied Electrodynamics, Russian
Academy of Sciences, Moscow, 125412 Russia}

\author{K.I. Kugel}
\affiliation{Institute for Theoretical and Applied Electrodynamics, Russian
Academy of Sciences, Moscow, 125412 Russia}
\affiliation{National Research University Higher School of Economics,
Moscow, 109028 Russia}

\author{Franco Nori}
\affiliation{Center for Emergent Matter Science, RIKEN, Wako-shi, Saitama,
351-0198, Japan}
\affiliation{Department of Physics, University of Michigan, Ann Arbor, MI
48109-1040, USA}

\begin{abstract}
Half-metallicity (full spin polarization of the Fermi surface) usually
occurs in strongly correlated electron systems. We demonstrate that doping
a spin-density wave insulator in the weak-coupling regime may also stabilize
half-metallic states. The undoped spin-density wave is formed by four
nested bands [i.e., each band is characterized by charge (electron/hole)
and spin (up/down) labels]. Of these four bands only two accumulate the
doped carriers, forming a half-metallic two-valley Fermi surface. Depending
on parameters, the spin polarizations of the electron-like and hole-like
valleys may be (i) parallel or (ii) antiparallel. The Fermi surface of (i)
is fully spin-polarized (similar to usual half-metals). Case (ii), referred
to as ``a spin-valley half-metal'', corresponds to complete polarization
with respect to the spin-valley operator. The properties of these states
are discussed.
\end{abstract}

\pacs{75.10.Lp,	
75.50.Ee,
75.50.Cc
}

\date{\today}

\maketitle

\textit{Introduction.} ---
Half-metallicity~\cite{first_half_met1983,half_met_review2008,hu2012half}
is a useful property for spintronics applications. Unlike usual metals,
which has both spin projections (spin-up and spin-down) on the Fermi
surface, half-metallicity implies that the electrons with only one spin
projection, for example, spin-up, reach the Fermi level. Spin-down states
are pushed away from the Fermi energy, making the ground state
non-invariant under spin-rotation transformations. A highly desirable
consequence~\cite{review_spintronics2004,hu2012half} of
half-metallicity is the perfect spin-polarization of electric current.
Experimental studies confirmed that a variety of real materials are
half-metals. For example, NiMnSb \cite{nimnsb_exp1990};
(La$_{0.7}$Sr$_{0.3}$)MnO$_{3}$~\cite{lasrmno_half_met_exp1998};
CrO$_2$~\cite{cro2_half_met_exp2001},
Co$_2$MnSi~\cite{co2mnsi_half_met_exp2014}, as well as many others.

From the theory standpoint, the half-metallicity of these compounds relies
on sizable electron-electron interactions, which are associated with
transition metal atoms. However, in recent years, the search for
`metal-free half-metals'
began~\cite{metal_free_hm2012,meta_free_hm2014}. Such systems could be
useful for bio-compatible applications, and, in general, are consistent
with current interest in carbon-based and organic-based mesoscopic 
systems~\cite{soriano2010,plastic_electr2010,Avouris2007,meso_review,
chinese_phys_silicene2014,bilayer_review2016}. 
It is difficult to expect a strong electron-electron interaction for
systems composed entirely of $s$- and $p$-elements. Thus, different
mechanisms for half-metallicity must be looked for.

\begin{figure}[t]
\centering
\includegraphics[width=0.99\columnwidth]{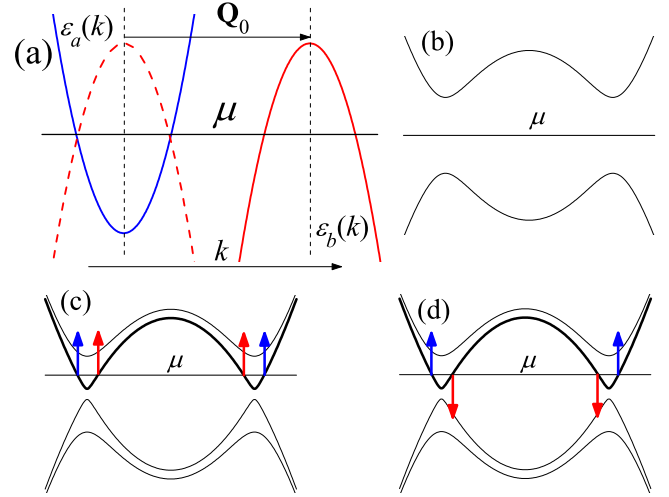}
\caption{(Color online) Schematics of the model electron energy bands and
spin structure: (a) electron bands in the absence of electron-hole
interaction and doping, (b)--(d) the electron-hole interaction is taken
into account; (b) the case of zero doping; (c) and (d) the doping is
non-zero, the spin polarization (arrows) of the Fermi surface sheets
corresponds to the CDW half-metal and to the spin-valley half-metal,
respectively (for details see text).
\label{valleys}
}
\end{figure}

In this paper, we discuss a novel possibility to generate a
half-metallicity. In brief, we demonstrate that doping a spin-density wave
(SDW) insulator may stabilize a type of half-metallic state. Let the
undoped system [see
Fig.~\ref{valleys}(a)]
have two nested Fermi surface sheets, which we will also refer to as
valleys. Let one sheet, or valley, correspond to electron states, and
another to hole states. Both valleys are spin-degenerate. The SDW
instability opens a gap generating an insulating ground state. We show
that, when doping is introduced, each valley becomes half-metallic. If the
spin polarizations of both sheets are parallel to each other,
Fig.~\ref{valleys}(c),
a half-metallic state, called below charge-density wave (CDW) half-metal,
emerges. For antiparallel polarizations, Fig.~\ref{valleys}(d), a
different half-metallic state, spin-valley half-metal, appears. The
properties of these two states are discussed below.

\textit{Model.} --- Our model has two bands, or valleys: an electronic band
$a$ and a hole band $b$, 
Fig.~\ref{valleys}(a), with the following
single-particle dispersion ($\hbar = 1$)
\begin{eqnarray}
\label{spectrum_a}
\varepsilon^a(\mathbf{k})=\frac{\mathbf{k}^2}{2m_a} +
\varepsilon^{a}_{\textrm{min}}-\mu, \quad \varepsilon^{a}_{\textrm{min}}<\varepsilon^a<\varepsilon^{a}_{\textrm{max}}, \\
\label{spectrum_b}
\varepsilon^b(\mathbf{k}+\mathbf{Q}_0) =
-\frac{\mathbf{k}^2}{2m_b}+\varepsilon^{b}_{\textrm{max}}-\mu, \quad \varepsilon^{b}_{\textrm{min}} < \varepsilon^b<\varepsilon^{b}_{\textrm{max}}.
\end{eqnarray}
Here band $a$ is centered at
${\bf k} = 0$,
and band $b$ at some finite momentum
${\bf Q}_0$.
Below, to simplify calculations, we will assume perfect electron-hole
symmetry:
$m_a = m_b = m$,
and
$\varepsilon^{b}_{\textrm{max}}=-\varepsilon^{a}_{\textrm{min}} =
\varepsilon_{\rm F}$,
consequently,
$\varepsilon^a ({\bf k})
=
-\varepsilon^b ({\bf k} + {\bf Q}_0)
=
\varepsilon_{\bf k}$.
Zero doping corresponds to
$\mu = 0$.
Undoped Fermi surface sheets for the $a$ and $b$ bands are characterized by
a single Fermi momentum
$k_{\rm F} = \sqrt{2m \varepsilon_{\rm F}}$,
and density of states (per spin projection)
$N_{\rm F} = {m k_{\rm F}}/(2\pi^2)$
at the Fermi energy. This provides a perfect nesting: a
translation of the electron Fermi surface by the vector $\mathbf{Q}_0$
completely superpose the sheets. The total Hamiltonian is equal to
\begin{equation}\label{ham_summa}
\hat{H}=\hat{H}_e+\hat{H}_{\textrm{int}}\,,
\end{equation}
where $\hat{H}_e$ is the single-electron term, described by the
dispersions~(\ref{spectrum_a}, \ref{spectrum_b}),
while $\hat{H}_{\textrm{int}}$
corresponds to the interaction between quasiparticles.

To treat the SDW instability, it is sufficient to keep in
$H_{\rm int}$
only the interaction between electrons in the $a$ and $b$ bands.  We also
assume that the interaction is a short-range one. We will focus mostly on
the following interaction term
\begin{equation}\label{eq::ham_int}
\hat{H}_{\textrm{int}}\!=\!g \int\!{d^3x\,\sum_{\sigma\sigma'}	
\psi^\dag_{a\sigma}\!(\mathbf{x})	\,
\psi_{a\sigma}\!(\mathbf{x})\,
\psi^\dag_{b\sigma'}\!(\mathbf{x})\,
\psi_{b\sigma'}\!(\mathbf{x})}\,.
\end{equation}
The effects of the neglected term will be discussed later.
In
Eq.~(\ref{eq::ham_int})
symbol
$\psi_{\alpha \sigma}$ denotes the usual fermionic field operator for band
$\alpha$ and spin $\sigma$. The coupling constant $g$ is positive, which
corresponds to repulsion, and weak
$g N_{\rm F} \ll 1$.

\textit{Spin-valley half-metal.} --- When the Fermi surface sheets of the
holes and the electrons perfectly match each other, model~(\ref{ham_summa})
describes spontaneous formation of the SDW or CDW orders. We start with the
SDW. The SDW ground state is believed to be unique (up to rotation of the
spin polarization axis), and well-described by a BCS-like mean-field
theory. The electron operators can be grouped into two sectors, labeled by
the index $\sigma=\pm 1$: sector $\sigma$ consists of $\psi_{a \sigma}$ and $\psi_{b {\bar \sigma}}$ (here ${\bar \sigma}$ means $-\sigma$). In the mean-field approach the sectors are decoupled and the SDW order parameter can be written as
\begin{equation}\label{rice}
\Delta_\sigma = \frac{g}{V} \sum_{\mathbf{k}}\left\langle
\psi^\dag_{\mathbf{k}a\sigma} \, 
\psi^{\phantom{\dag}}_{\mathbf{k}b\bar{\sigma}}\right\rangle\,,
\end{equation}
where $V$ is the system volume, and
$\langle \ldots \rangle$
denotes the diagonal matrix element for the ground state. At zero doping,
the sectors are degenerate: $\Delta_\uparrow = \Delta_\downarrow =
\Delta_0$, where $\Delta_0$ is the order parameter at perfect nesting. This
equality implies that the SDW polarization in real space is along the $x$-axis:
\begin{eqnarray}\label{eq::sx}
\langle S^x(\mathbf{r})\rangle&=&\frac{\Delta_\uparrow + \Delta_\downarrow}{g}\cos ({\bf Q}_0 {\bf r})=\frac{2\Delta_0}{g}\cos({\bf Q}_0 {\bf r})\,,\\
\label{eq::sy}
\langle S^y(\mathbf{r})\rangle&=&\frac{\Delta_\uparrow - \Delta_\downarrow}{2g}\sin({\bf Q}_0 {\bf r})\equiv 0\,.
\end{eqnarray}
For $\Delta_0$ one can write
$\Delta_0 \approx \varepsilon_{\rm F} \exp \left( -1/g N_{\rm F} \right)$.

Doping destroys the perfect Fermi surface nesting and the number of low-energy states competing to become the true ground state sharply grows. Both incommensurate and inhomogeneous phases~\cite{Rice,tokatly1992,eremin_chub2010,PrlOur,PrbOur,PrbROur,Sboychakov_PRB2013_PS_pnict,PrbVOur,prb_sl2017}
were considered for Hamiltonian~(\ref{ham_summa}) and its numerous variations. Here we argue that the half-metallic state is yet another viable contender in the case of imperfect nesting.

The grand potential of our system $\Omega$ at zero temperature and finite doping $x$ is a sum of two partial grand potentials $\Omega = \sum_\sigma \Omega_\sigma$, where
\begin{eqnarray}\label{eq::grand_pot}
\frac{\Omega_\sigma}{V}
&=&
\frac{\Delta_\sigma^2}{g}
-
\frac{1}{V}
\sum_{\bf k}
	\left[
		\left(
			\mu+\sqrt{\varepsilon_{\bf k}^2+\Delta_\sigma^2}
		\right)
		\right.
\\
\nonumber
&+&
		\left.
		\left(
			\mu-\sqrt{\varepsilon_{\bf k}^2+\Delta_\sigma^2}
		\right)
		\theta
		\left(
			\mu-\sqrt{\varepsilon_{\bf k}^2+\Delta_\sigma^2}
		\right)
	\right],
\end{eqnarray}
and $\theta (z)$ is the Heaviside step-function.
To describe doping it is convenient to introduce $x_\sigma = - \partial
\Omega_\sigma/\partial \mu$. The physical meaning of partial doping
$x_\sigma$ is the amount of doped charge accumulated in sector $\sigma$. The value of $\Delta_{\sigma}$ corresponds to the minimum of $\Omega_\sigma$. Thus, one has to solve the system of equations
\begin{eqnarray}\label{eq::MFparam}
\frac{\partial \Omega_\sigma}{\partial \Delta_\sigma} = 0,
\qquad
x_\uparrow + x_\downarrow = x,
\end{eqnarray}
to determine $\mu$ and $\Delta_\sigma$. 
Equations~(\ref{eq::grand_pot},~\ref{eq::MFparam}) are valid provided that (a) the state remains homogeneous, and (b) the SDW order parameter remains commensurate even at finite doping.

Since the two sectors $\sigma$ are decoupled, one can calculate
$\Delta_\sigma$ and $\mu$ as functions of $x_\sigma$. The result is~\cite{tokatly1992,our_chrom2013,prb_sl2017}
\begin{eqnarray}\label{eq::doped_sdw_delta}
\Delta_\sigma
=
\Delta_0 \sqrt{1 - \frac{x_\sigma}{N_{\rm F} \Delta_0}},
\quad
\text{and}
\quad
\mu = \Delta_0 - \frac{x_\sigma}{2 N_{\rm F}}.
\end{eqnarray}
Thus, the order parameter decreases at finite doping, and the homogeneous
commensurate state becomes completely unstable for
$x_\sigma > x_c = N_{\rm F} \Delta_0$.

It is often implicitly assumed (see, e.g,
Refs.~\cite{Rice,our_chrom2013,prb_sl2017}) that the charge carriers are
spread evenly between both sectors of $\sigma$, and the degeneracy of
$\Delta_\sigma$ persists even for finite $x$. Yet, it is easy to show that
the spontaneous lifting of this degeneracy optimizes the energy. To prove
this, consider the system's free energy 
$F=\sum_\sigma F_\sigma$,
where the partial free energy
$F_\sigma = \Omega_\sigma+\mu x_\sigma$
can be calculated as
$F_\sigma (x_\sigma) = F_\sigma (0) +
\int_0^{x_\sigma} d{\bar x} \mu ({\bar x})$.
One thus obtains
\begin{eqnarray}\label{eq::Ftotal}
\frac{F}{V}
=
\sum_\sigma \frac{F_\sigma}{V}
=-
N_{\rm F} \Delta_0^2 + \Delta_0 x
-
\frac{x_\uparrow^2 + x_\downarrow^2}{4 N_{\rm F}},
\end{eqnarray}
where we took into account that
$F_\sigma (0) = - N_{\rm F} \Delta^2_0/2$.
Note that only the third term in Eq.~(\ref{eq::Ftotal}) depends on the distribution of the dopants among the two sectors. It is easy to check that, if
$x_\sigma = x$
and
$x_{\bar \sigma} = 0$,
the third term, together with $F$, is the smallest. In other words, for
fixed $x$, the most stable spatially-homogeneous state of the model
corresponds to the case when all the doped charge accumulates in a given
sector. The other sector is completely free of the extra charge carriers.
Therefore
\begin{eqnarray}\label{eq::F_half}
\frac{F}{V}
=
-
N_{\rm F} \Delta_0^2 + \Delta_0 x
-
\frac{x^2}{4 N_{\rm F}},
\\
\label{eq::mu_half}
\mu = \Delta_0 - \frac{x}{2N_{\rm F}},
\\
\label{eq::delta_half}
\Delta_\sigma (x)
=
\Delta_0 \sqrt{1 - \frac{x}{N_{\rm F} \Delta_0}},
\quad
\Delta_{\bar \sigma} (x) = \Delta_0.
\end{eqnarray}
These relations are valid for not too strong doping
$x>N_{\rm F} \Delta_0$.
An important feature of Eq.~(\ref{eq::F_half}) is that the second
derivative $\partial^2 F/\partial x^2$ is negative. This means that the
doped system is unstable with respect to electronic phase separation~\cite{tokatly1992,PrbROur,Sboychakov_PRB2013_PS_pnict,our_chrom2013,IrkhinPRB2010}. However, the Coulomb interaction can suppress the phase separation~\cite{di_castro1,bianconi2015intrinsic}. Thus, it is reasonable to study the properties of the homogeneous state.

It follows from Eqs.~(\ref{eq::mu_half},~\ref{eq::delta_half}) that
$\Delta_\sigma (x) < \mu (x) < \Delta_{\bar \sigma} (x) = \Delta_0$,
when
$x>0$.
This means that in the sector $\bar \sigma$, the order parameter
remains equal to $\Delta_0$ and, since the chemical potential is lower than
$\Delta_{\bar \sigma}$, no charge enters this sector, see
Fig.~\ref{valleys}(d). In the sector $\sigma$, two Fermi surface sheets emerge. They are fixed by the equation $\varepsilon^2_{\bf k} = [\mu (x)]^2 - [\Delta_\sigma(x)]^2$, which is equivalent to
\begin{eqnarray}
\varepsilon_{\bf k}
=
\pm \frac{x}{2N_{\rm F}}.
\end{eqnarray}
As the doped charges are distributed unevenly between the sectors, the
doped state acquires non-trivial macroscopic quantum numbers. To
characterize the macroscopic state, it is useful to specify spin ($\hat S$) and spin-valley ($\hat S_{\rm v}$) operators
\begin{eqnarray}
{\hat S} = \sum_{\alpha\sigma} \sigma {\hat N}_{\alpha\sigma},
\quad
{\hat S}_{\rm v}
=
\sum_{\alpha\sigma}
	\sigma v_\alpha {\hat N}_{\alpha \sigma}\,,
\end{eqnarray}
where $\alpha=a,b$, and the valley index
$v_\alpha$
is defined as follows:
$v_a = 1$ and $v_b = -1$. The operator ${\hat N}_{\alpha\sigma}$ counts the number of electrons with spin $\sigma$ in valley $\alpha$
\begin{eqnarray}
{\hat N}_{\alpha\sigma}
=
\sum_{\bf k}
	\psi^\dag_{{\bf k}\alpha \sigma}\,
	\psi^{\vphantom{\dagger}}_{{\bf k}\alpha\sigma}.
\end{eqnarray}
The Hamiltonian~(\ref{ham_summa}) commutes with both ${\hat S}$ and
${\hat S}_{\rm v}$. The field operators satisfy obvious commutation rules
\begin{eqnarray}
\left[ {\hat S}, \psi_{\alpha \sigma} \right]
=
\sigma \psi_{\alpha \sigma},
\quad
\left[ {\hat S}_{\rm v}, \psi_{\alpha \sigma} \right]
=
\sigma v_\alpha \psi_{\alpha \sigma}.
\end{eqnarray}
In other words, in addition to the spin quantum number $\sigma$, a field $\psi_{\alpha\sigma}$ can be characterized by the spin-valley projection $\sigma v_\alpha$.

It is easy to check that in the sector $\sigma$ both $\psi_{a \sigma}$ and
$\psi_{b {\bar\sigma}}$ carry the same spin-valley quantum equal to
$+\sigma$. In the sector ${\bar \sigma}$, the field operators correspond to
a ${-\sigma}$ quantum of ${\hat S}_{\rm v}$. That is, the Fermi surface of
the doped system is characterized by the single projection of the
spin-valley operator. The Fermi surface sheets with the opposite projection
of ${\hat S}_{\rm v}$ are absent, since the sector ${\bar \sigma}$ is
gapped. Thus, the doped system can be referred to as 
\textit{a spin-valley half-metal}:
like a classical half-metal, our system exhibits complete polarization of
the Fermi surface; however, in contrast to the usual half-metal, the
polarization is not the spin polarization, but rather, the spin-valley one.
Therefore, the electrical current through the spin-valley half-metal is
completely spin-valley polarized.

Since the sector
${\bar \sigma}$
is free from doped electrons, the average values of $\hat N_{a{\bar
\sigma}}$ and $\hat N_{b {\sigma}}$ remain unaffected by the doping, while
$\langle \hat N_{a{\sigma}} \rangle$ and $\langle \hat N_{b {\bar \sigma}}
\rangle$ change. Taking the average occupation numbers $N_{\alpha \sigma} =
\langle {\hat N}_{\alpha \sigma} \rangle$ in the undoped state to be zero,
we can write $N_{a {\bar \sigma}}=N_{b {\sigma}}=0$, and $N_{a
{\sigma}}+N_{b {\bar \sigma}}=xV$. Consequently, $S_{\rm v} = \langle {\hat
S}_{\rm v} \rangle$ is proportional to $x$. Namely, $S_{\rm v} = \sigma x
V$. In a system with the perfect electron-hole symmetry we have $N_{a
{\sigma}} = N_{b {\bar \sigma}} = x V/2$, which corresponds to $S = \langle
{\hat S} \rangle \equiv 0$,
for any $x$. When the symmetry is absent: $|S| \propto x$. However, the net
spin polarization of the spin-valley half-metal satisfies the inequality 
$|S| < |S_{\rm v}|$.

Doping also affects the SDW order inherited from the undoped state.
Intuitively, since the charge enters only one of the two sectors, the
symmetry between sectors $\sigma$ disappears for $x>0$.
[Eqs.~(\ref{eq::delta_half}) prove this.] Thus, simple SDW is replaced by a
more complicated order parameter: analyzing 
Eqs.~(\ref{eq::sx},~\ref{eq::sy}) one can prove that, at finite doping, a
circularly-polarized spin component emerges
\begin{equation}\label{eq::sdw_circ}
\{\delta S^x(\mathbf{r}),\delta S^y(\mathbf{r}) \}\!\propto\!(\Delta_\uparrow - \Delta_\downarrow )\{\cos ({\bf Q}_0 {\bf r}),\sin({\bf Q}_0 {\bf r})\}.
\end{equation}
The amplitude of this component increases when $x$ grows.

\textit{From spin-valley half-metal to CDW half-metal.} --- In addition to the
expected invariance with respect to simultaneous rotations of all fermion
spins, our model Hamiltonian allows for a broader class of
symmetries: it remains unchanged, even if the electron and hole spins are
transformed by two different rotation operators. This observation can be
trivially proven in the absence of interaction ($g=0$). In the case of a
generic interaction, this symmetry does not apply. However, if the
interaction is short-range, as in Eq.~(\ref{eq::ham_int}), the invariance of the
Hamiltonian under such transformations remains. Indeed, the integrand in Eq.~(\ref{eq::ham_int}) is $\propto \rho_{\rm e} \rho_{\rm h}$, where $\rho_{\rm e}$ and $\rho_{\rm h}$ are the density operators for electrons and holes, which both are invariant under separate rotations of the electron and hole spins. Therefore, the substitution
\begin{eqnarray}\label{eq::substit}
\psi_{b \uparrow}
\rightarrow
\psi_{b \downarrow},
\quad
\psi_{b {\downarrow}}
\rightarrow
\psi_{b \uparrow}
\end{eqnarray}
corresponds to a symmetry of the model. Consequently, Eq.~(\ref{eq::substit})
either preserves the ground state, or transforms one ground state into
another one. Since the order parameter, Eq.~(\ref{rice}), changes under the
transformation~(\ref{eq::substit}), we must conclude that a new ground
state is generated by such a substitution. If we start with the spin-valley half-metal ground state, what kind of new state the transformation~(\ref{eq::substit}) brings us?

Consider the SDW polarization, Eq.~(\ref{eq::sx}), at zero doping. Under
the transformation~(\ref{eq::substit}) the SDW is replaced by charge-density wave  with finite average value of the density operator ${\hat \rho}_{{\bf Q}_0}$:
\begin{eqnarray}
\langle {\hat S}^x_{{\bf Q}_0} \rangle
\!=\!
\sum_\sigma
	\langle
		\psi^\dag_{\mathbf{k}a\sigma}
		\psi^{\phantom{\dag}}_{\mathbf{k}b\bar{\sigma}}
	\rangle
\rightarrow
\sum_\sigma
	\langle
		\psi^\dag_{\mathbf{k}a\sigma}
		\psi^{\phantom{\dag}}_{\mathbf{k}b\sigma}
	\rangle
\!=\!
\langle{\hat \rho}_{{\bf Q}_0}\rangle.
\end{eqnarray}

Calculations identical (up to relabeling) to the case of the SDW order
demonstrate that for $x > 0$ the charge carriers accumulate in a
single mean-field sector. However, the sector composition is changed by
the transformation~(\ref{eq::substit}): sector $\sigma$
consists of $\psi_{a \sigma}$ and $\psi_{b \sigma}$. Unlike the case of
spin-valley half-metals, now both electronic fields within a single sector have
the same spin projection. Therefore, if the doped charge enters sector
$\sigma$, both Fermi surface sheets have identical spin polarizations equal
to $\sigma$, see Fig.~\ref{valleys}(c). This perfect polarization of the
Fermi surface is a hallmark feature of half-metals. Thus, the spin-valley
half-metal is related to the CDW half-metal by the
substitution~(\ref{eq::substit}), and both states are degenerate within our
model. This connection becomes apparent if we notice 
that~(\ref{eq::substit}) 
switches the operators $\hat S$ and $\hat S_{\rm v}$. Consequently, in the
CDW half-metal
$S=\sigma x V$ and $|S_{\rm v}| < |S|$.
When $x>0$, in addition to the CDW order parameter, the SDW order parameter
$\langle S_{{\bf Q}_0}^z \rangle$ is generated. It grows monotonously as
$x$ grows. This is a direct analog of
Eq.~(\ref{eq::sdw_circ}).

Note, however, that the degeneracy between the SDW and CDW ground states is
an artifact of the short-range interaction, Eq.~(\ref{eq::ham_int}), which
possesses extra symmetries absent in more realistic models. The effects of
more generic interaction operators are discussed below.

\textit{Discussion.} --- The interaction Eq.~(\ref{eq::ham_int}) is not the
most general form of electron-electron coupling. In particular, the
``exchange" term
\begin{equation}
\label{eq::ham_ex}
\hat{H}_{\textrm{ex}}\!
=\!
g_\perp \int\!{d^3x\,
\sum_{\sigma\sigma'}
	\psi^\dag_{a\sigma} \psi_{b\sigma}^{\vphantom{\dagger}}
	\psi^\dag_{b\sigma'} \psi_{a\sigma'}^{\vphantom{\dagger}}
      }
\end{equation}
should be accounted for. The coupling constant $g_\perp > 0$ describes a
repulsive interaction at finite momentum ${\bf Q}_0$. The ``exchange"
term~(\ref{eq::ham_ex})
immediately lifts the degeneracy between the SDW and CDW, in favor of SDW.
This means that, for finite doping, the spin-valley half-metal is more
stable than the CDW half-metal. Also, other factors could favor the CDW
half-metal, for example, the proximity to a lattice instability. An
external magnetic field acts similarly, since the total spin of the CDW
half-metal exceeds the spin of the spin-valley half-metal.

\begin{figure}[t]
\centering
\includegraphics[width=0.99\columnwidth]{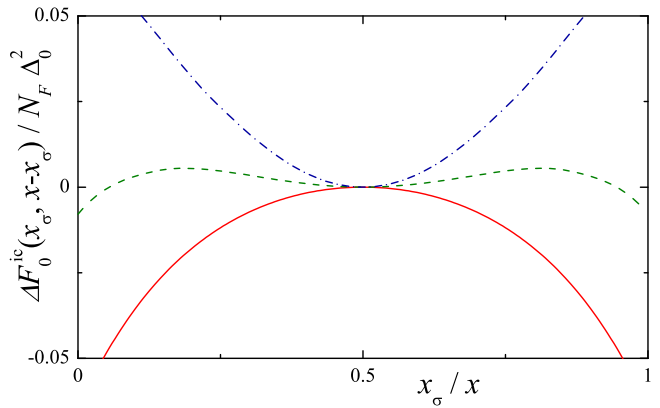}
\caption{\label{FigFreeEnergy} (Color online) Dependence of $\Delta F_0^{\rm ic}(x_\sigma,x-x_\sigma)\equiv F_0^{\rm ic}(x_\sigma)+F_0^{\rm ic}(x-x_\sigma)-2F_0^{\rm ic}(x/2) $ on partial doping $x_\sigma$, calculated at $T=0$ and fixed total doping $x=1.4N_F\Delta_0$ [(red) solid curve], $x=1.76N_F\Delta_0$ [(green) dashed curve], and $x=2.0N_F\Delta_0$ [(blue) dot-dashed curve].}
\end{figure}

We assumed that the Coulomb interaction guarantees the homogeneity of the
electron liquid. Thus, in the above discussion we neglected the possibility
of phase separation. In addition, the incommensurate SDW states were not
considered. While the detailed study of such states is an interesting goal
for future research, we do not expect that this modification would affect
significantly the stability of the half-metallic phases, at least at some
doping region. Indeed, at the mean-field level the free energy in the
presence of the incommensurate SDW equals
\begin{eqnarray}
F^{\rm ic} (x) = \min_{x_\uparrow + x_\downarrow = x}
\left[	F_0^{\rm ic} (x_\uparrow) + F_0^{\rm ic} (x_\downarrow)\right]\,,
\end{eqnarray}
where $F_0^{\rm ic} (x_\sigma)$ is the free energy of a sector with partial
doping $x_\sigma$. As above, the free energy of the system is found by
minimization under the condition 
$x_\uparrow + x_\downarrow = x$. 
We calculated $F_0^{\rm ic}(x_\sigma)$ numerically, as described in
Ref.~\cite{our_chrom2013}. Our analysis shows that $\partial^2 F_0^{\rm
ic}(x_\sigma) / \partial x_\sigma^2 < 0$ for $x_\sigma$ less than the threshold value $x^*\cong0.83N_F\Delta_0$. This is a rather general feature of a system with imperfect nesting~\cite{tokatly1992,PrbROur,Sboychakov_PRB2013_PS_pnict,our_chrom2013,IrkhinPRB2010}. Since the second derivative of
$F_0^{\rm ic}$ is negative, the sum $F_0^{\rm ic} (x_\uparrow) + F_0^{\rm
ic} (x-x_\uparrow)$ as a function of $x_\uparrow \in [0, x]$ is concave.
Consequently, the extremum of the latter sum at $x_\uparrow = x/2$
corresponds to a maximum, not a minimum (see Fig.~\ref{FigFreeEnergy}). Thus, the total free energy is minimized as follows
\begin{eqnarray}
F^{\rm ic} (x)=F_0^{\rm ic} (x) + F_0^{\rm ic} (0),\text{\ \ at\ \ }x_\sigma = x,
\quad
x_{\bar \sigma} = 0\,.
\end{eqnarray}
Undoped sector ${\bar \sigma}$ remains insulating. All doped charge goes to
sector $\sigma$, which becomes metallic, with a well-defined Fermi surface.
Thus, we recover the spin-valley half-metal with an incommensurate SDW.

If $x_\sigma>x^*$,
then
$\partial^2 F_0^{\rm ic}(x_\sigma)/\partial x_\sigma^2 > 0$,
and the total free energy  $F_0^{\rm
ic}(x_\sigma)+F_0^{\rm ic}(x-x_\sigma)$ acquires a local minimum at
$x_{\uparrow}=x_{\downarrow}=x/2$ (see Fig.~\ref{FigFreeEnergy}). When
doping increases even further, this minimum becomes a global minimum for
$x\cong1.8N_F\Delta_0$. Thus, the first order transition from
incommensurate spin-valley half-metal to common incommensurate SDW phase occurs at this point.

We assume that both the electron and hole sheets in the Fermi surface have
a spherical shape and are perfectly nested at zero doping. If these
shapes are different, the range of doping where
$\partial^2 F^{\rm ic}_0 (x)/\partial x^2 < 0$
diminishes~\cite{Sboychakov_PRB2013_PS_pnict}. When the geometry of the
sheets differs significantly, one has
$\partial^2 F^{\rm ic}_0 (x)/\partial x^2 > 0$
for all $x$, and the half-metal states become impossible.

To conclude, we demonstrated that doping a SDW state with perfectly-nested
Fermi surface sheets stabilizes a half-metal-like ground state. Depending
on microscopic parameters and the external magnetic field, such ground
state could be either CDW half-metal with complete spin-polarization of the
Fermi surface or spin-valley half-metal. The Fermi surface of the latter
state is characterized by a perfect polarization in the spin-valley space.
While the CDW half-metal supports purely spin-polarized currents, which is
a natural consequence of the Fermi surface polarization, the spin-valley
half-metal supports spin-valley-polarized currents.

\textit{Acknowledgments.} --- This work is partially supported by the
Russian Foundation for Basic Research (Projects 17-02-00323 and
15-02-02128), JSPS-RFBR Joint project No 17-52-50023, RIKEN iTHES Project,
the MURI Center for Dynamic Magneto-Optics via the AFOSR award No.
FA9550-14-1-0040, the IMPACT program of JST, a Grant-in-Aid for Scientific
Research (A), CREST, and a grant from the John Templeton Foundation.


\begin{thebibliography}{29}
\expandafter\ifx\csname natexlab\endcsname\relax\def\natexlab#1{#1}\fi
\expandafter\ifx\csname bibnamefont\endcsname\relax
  \def\bibnamefont#1{#1}\fi
\expandafter\ifx\csname bibfnamefont\endcsname\relax
  \def\bibfnamefont#1{#1}\fi
\expandafter\ifx\csname citenamefont\endcsname\relax
  \def\citenamefont#1{#1}\fi

\bibitem[{\citenamefont{de~Groot et~al.}(1983)\citenamefont{de~Groot, Mueller,
  van Engen, and Buschow}}]{first_half_met1983}
\bibinfo{author}{\bibfnamefont{R.A.} \bibnamefont{de~Groot}},
  \bibinfo{author}{\bibfnamefont{F.M.} \bibnamefont{Mueller}},
  \bibinfo{author}{\bibfnamefont{P.G.} \bibnamefont{van Engen}},
  \bibnamefont{and} \bibinfo{author}{\bibfnamefont{K.H.J.}
  \bibnamefont{Buschow}}, {``}\bibinfo{title}{New Class of Materials:
  Half-Metallic Ferromagnets},{''} \bibinfo{journal}{Phys. Rev. Lett.}
  \textbf{\bibinfo{volume}{50}}, \bibinfo{pages}{2024} (\bibinfo{year}{1983}).

\bibitem[{\citenamefont{Katsnelson et~al.}(2008)\citenamefont{Katsnelson,
  Irkhin, Chioncel, Lichtenstein, and de~Groot}}]{half_met_review2008}
\bibinfo{author}{\bibfnamefont{M.I.} \bibnamefont{Katsnelson}},
  \bibinfo{author}{\bibfnamefont{V.Y.} \bibnamefont{Irkhin}},
  \bibinfo{author}{\bibfnamefont{L.}~\bibnamefont{Chioncel}},
  \bibinfo{author}{\bibfnamefont{A.I.} \bibnamefont{Lichtenstein}},
  \bibnamefont{and} \bibinfo{author}{\bibfnamefont{R.A.}
  \bibnamefont{de~Groot}}, {``}\bibinfo{title}{Half-metallic ferromagnets: From
  band structure to many-body effects},{''} \bibinfo{journal}{Rev. Mod. Phys.}
  \textbf{\bibinfo{volume}{80}}, \bibinfo{pages}{315} (\bibinfo{year}{2008}).

\bibitem[{\citenamefont{Hu}(2012)}]{hu2012half}
\bibinfo{author}{\bibfnamefont{X.}~\bibnamefont{Hu}},
  {``}\bibinfo{title}{Half-Metallic Antiferromagnet as a Prospective Material
  for Spintronics},{''} \bibinfo{journal}{Adv. Mater.}
  \textbf{\bibinfo{volume}{24}}, \bibinfo{pages}{294} (\bibinfo{year}{2012}).

\bibitem[{\citenamefont{{\v{Z}}uti{\'{c}}
  et~al.}(2004)\citenamefont{{\v{Z}}uti{\'{c}}, Fabian, and
  Das~Sarma}}]{review_spintronics2004}
\bibinfo{author}{\bibfnamefont{I.}~\bibnamefont{{\v{Z}}uti{\'{c}}}},
  \bibinfo{author}{\bibfnamefont{J.}~\bibnamefont{Fabian}}, \bibnamefont{and}
  \bibinfo{author}{\bibfnamefont{S.}~\bibnamefont{Das~Sarma}},
  {``}\bibinfo{title}{Spintronics: Fundamentals and applications},{''}
  \bibinfo{journal}{Rev. Mod. Phys.} \textbf{\bibinfo{volume}{76}},
  \bibinfo{pages}{323} (\bibinfo{year}{2004}).

\bibitem[{\citenamefont{Hanssen et~al.}(1990)\citenamefont{Hanssen, Mijnarends,
  Rabou, and Buschow}}]{nimnsb_exp1990}
\bibinfo{author}{\bibfnamefont{K.E.H.M.} \bibnamefont{Hanssen}},
  \bibinfo{author}{\bibfnamefont{P.E.} \bibnamefont{Mijnarends}},
  \bibinfo{author}{\bibfnamefont{L.P.L.M.} \bibnamefont{Rabou}},
  \bibnamefont{and} \bibinfo{author}{\bibfnamefont{K.H.J.}
  \bibnamefont{Buschow}}, {``}\bibinfo{title}{Positron-annihilation study of
  the half-metallic ferromagnet NiMnSb: Experiment},{''}
  \bibinfo{journal}{Phys. Rev. B} \textbf{\bibinfo{volume}{42}},
  \bibinfo{pages}{1533} (\bibinfo{year}{1990}).

\bibitem[{\citenamefont{Park et~al.}(1998)\citenamefont{Park, Vescovo, Kim,
  Kwon, Ramesh, and Venkatesan}}]{lasrmno_half_met_exp1998}
\bibinfo{author}{\bibfnamefont{J.-H.} \bibnamefont{Park}},
  \bibinfo{author}{\bibfnamefont{E.}~\bibnamefont{Vescovo}},
  \bibinfo{author}{\bibfnamefont{H.-J.} \bibnamefont{Kim}},
  \bibinfo{author}{\bibfnamefont{C.}~\bibnamefont{Kwon}},
  \bibinfo{author}{\bibfnamefont{R.}~\bibnamefont{Ramesh}}, \bibnamefont{and}
  \bibinfo{author}{\bibfnamefont{T.}~\bibnamefont{Venkatesan}},
  {``}\bibinfo{title}{Direct evidence for a half-metallic ferromagnet},{''}
  \bibinfo{journal}{Nature} \textbf{\bibinfo{volume}{392}},
  \bibinfo{pages}{794} (\bibinfo{year}{1998}).

\bibitem[{\citenamefont{Ji et~al.}(2001)\citenamefont{Ji, Strijkers, Yang,
  Chien, Byers, Anguelouch, Xiao, and Gupta}}]{cro2_half_met_exp2001}
\bibinfo{author}{\bibfnamefont{Y.}~\bibnamefont{Ji}},
  \bibinfo{author}{\bibfnamefont{G.J.} \bibnamefont{Strijkers}},
  \bibinfo{author}{\bibfnamefont{F.Y.} \bibnamefont{Yang}},
  \bibinfo{author}{\bibfnamefont{C.L.} \bibnamefont{Chien}},
  \bibinfo{author}{\bibfnamefont{J.M.} \bibnamefont{Byers}},
  \bibinfo{author}{\bibfnamefont{A.}~\bibnamefont{Anguelouch}},
  \bibinfo{author}{\bibfnamefont{G.}~\bibnamefont{Xiao}}, \bibnamefont{and}
  \bibinfo{author}{\bibfnamefont{A.}~\bibnamefont{Gupta}},
  {``}\bibinfo{title}{Determination of the Spin Polarization of Half-Metallic
  ${\mathrm{CrO}}_{2}$ by Point Contact Andreev Reflection},{''}
  \bibinfo{journal}{Phys. Rev. Lett.} \textbf{\bibinfo{volume}{86}},
  \bibinfo{pages}{5585} (\bibinfo{year}{2001}).

\bibitem[{\citenamefont{Jourdan et~al.}(2014)\citenamefont{Jourdan,
  Min{\'{a}}¡r, Braun, Kronenberg, Chadov, Balke, Gloskovskii, Kolbe, Elmers,
  Sch{\"{o}}nhense et~al.}}]{co2mnsi_half_met_exp2014}
\bibinfo{author}{\bibfnamefont{M.}~\bibnamefont{Jourdan}},
  \bibinfo{author}{\bibfnamefont{J.}~\bibnamefont{Min{\'{a}}¡r}},
  \bibinfo{author}{\bibfnamefont{J.}~\bibnamefont{Braun}},
  \bibinfo{author}{\bibfnamefont{A.}~\bibnamefont{Kronenberg}},
  \bibinfo{author}{\bibfnamefont{S.}~\bibnamefont{Chadov}},
  \bibinfo{author}{\bibfnamefont{B.}~\bibnamefont{Balke}},
  \bibinfo{author}{\bibfnamefont{A.}~\bibnamefont{Gloskovskii}},
  \bibinfo{author}{\bibfnamefont{M.}~\bibnamefont{Kolbe}},
  \bibinfo{author}{\bibfnamefont{H.}~\bibnamefont{Elmers}},
  \bibinfo{author}{\bibfnamefont{G.}~\bibnamefont{Sch{\"{o}}nhense}},
  \bibnamefont{et~al.}, {``}\bibinfo{title}{Direct observation of
  half-metallicity in the Heusler compound Co{$_2$}MnSi},{''}
  \bibinfo{journal}{Nat. Commun.} \textbf{\bibinfo{volume}{5}},
  \bibinfo{pages}{3974} (\bibinfo{year}{2014}).

\bibitem[{\citenamefont{Du et~al.}(2012)\citenamefont{Du, Sanvito, and
  Smith}}]{metal_free_hm2012}
\bibinfo{author}{\bibfnamefont{A.}~\bibnamefont{Du}},
  \bibinfo{author}{\bibfnamefont{S.}~\bibnamefont{Sanvito}}, \bibnamefont{and}
  \bibinfo{author}{\bibfnamefont{S.C.} \bibnamefont{Smith}},
  {``}\bibinfo{title}{First-Principles Prediction of Metal-Free Magnetism and
  Intrinsic Half-Metallicity in Graphitic Carbon Nitride},{''}
  \bibinfo{journal}{Phys. Rev. Lett.} \textbf{\bibinfo{volume}{108}},
  \bibinfo{pages}{197207} (\bibinfo{year}{2012}).

\bibitem[{\citenamefont{Hashmi and Hong}(2014)}]{meta_free_hm2014}
\bibinfo{author}{\bibfnamefont{A.}~\bibnamefont{Hashmi}} \bibnamefont{and}
  \bibinfo{author}{\bibfnamefont{J.}~\bibnamefont{Hong}},
  {``}\bibinfo{title}{Metal free half metallicity in 2D system: structural and
  magnetic properties of g-C$_4$N$_3$ on BN},{''} \bibinfo{journal}{Sci. Rep.}
  \textbf{\bibinfo{volume}{4}}, \bibinfo{pages}{4374} (\bibinfo{year}{2014}).

\bibitem[{\citenamefont{Soriano and Fern\'andez-Rossier}(2010)}]{soriano2010}
\bibinfo{author}{\bibfnamefont{D.}~\bibnamefont{Soriano}} \bibnamefont{and}
  \bibinfo{author}{\bibfnamefont{J.}~\bibnamefont{Fern\'andez-Rossier}},
  {``}\bibinfo{title}{Spontaneous persistent currents in a quantum spin Hall
  insulator},{''} \bibinfo{journal}{Phys. Rev. B}
  \textbf{\bibinfo{volume}{82}}, \bibinfo{pages}{161302}
  (\bibinfo{year}{2010}).

\bibitem[{\citenamefont{Klauk}(2010)}]{plastic_electr2010}
\bibinfo{author}{\bibfnamefont{H.}~\bibnamefont{Klauk}},
  {``}\bibinfo{title}{Organic thin-film transistors},{''}
  \bibinfo{journal}{Chem. Soc. Rev.} \textbf{\bibinfo{volume}{39}},
  \bibinfo{pages}{2643} (\bibinfo{year}{2010}).

\bibitem[{\citenamefont{Avouris et~al.}(2007)\citenamefont{Avouris, Chen, and
  Perebeinos}}]{Avouris2007}
\bibinfo{author}{\bibfnamefont{P.}~\bibnamefont{Avouris}},
  \bibinfo{author}{\bibfnamefont{Z.}~\bibnamefont{Chen}}, \bibnamefont{and}
  \bibinfo{author}{\bibfnamefont{V.}~\bibnamefont{Perebeinos}},
  {``}\bibinfo{title}{Carbon-based electronics},{''} \bibinfo{journal}{Nat.
  Nanotechnol.} \textbf{\bibinfo{volume}{2}}, \bibinfo{pages}{605}
  (\bibinfo{year}{2007}).

\bibitem[{\citenamefont{Rozhkov et~al.}(2011)\citenamefont{Rozhkov, Giavaras,
  Bliokh, Freilikher, and Nori}}]{meso_review}
\bibinfo{author}{\bibfnamefont{A.}~\bibnamefont{Rozhkov}},
  \bibinfo{author}{\bibfnamefont{G.}~\bibnamefont{Giavaras}},
  \bibinfo{author}{\bibfnamefont{Y.P.} \bibnamefont{Bliokh}},
  \bibinfo{author}{\bibfnamefont{V.}~\bibnamefont{Freilikher}},
  \bibnamefont{and} \bibinfo{author}{\bibfnamefont{F.}~\bibnamefont{Nori}},
  {``}\bibinfo{title}{Electronic properties of mesoscopic graphene structures:
  Charge confinement and control of spin and charge transport},{''}
  \bibinfo{journal}{Phys. Rep.} \textbf{\bibinfo{volume}{503}},
  \bibinfo{pages}{77 } (\bibinfo{year}{2011}).

\bibitem[{\citenamefont{Sa-Ke et~al.}(2014)\citenamefont{Sa-Ke, Hong-Yu,
  Yong-Hong, and Jun}}]{chinese_phys_silicene2014}
\bibinfo{author}{\bibfnamefont{W.}~\bibnamefont{Sa-Ke}},
  \bibinfo{author}{\bibfnamefont{T.}~\bibnamefont{Hong-Yu}},
  \bibinfo{author}{\bibfnamefont{Y.}~\bibnamefont{Yong-Hong}},
  \bibnamefont{and} \bibinfo{author}{\bibfnamefont{W.}~\bibnamefont{Jun}},
  {``}\bibinfo{title}{Spin and valley half metal induced by staggered potential
  and magnetization in silicene},{''} \bibinfo{journal}{Chin. Phys. B}
  \textbf{\bibinfo{volume}{23}}, \bibinfo{pages}{017203}
  (\bibinfo{year}{2014}).

\bibitem[{\citenamefont{Rozhkov et~al.}(2016)\citenamefont{Rozhkov, Sboychakov,
  Rakhmanov, and Nori}}]{bilayer_review2016}
\bibinfo{author}{\bibfnamefont{A.}~\bibnamefont{Rozhkov}},
  \bibinfo{author}{\bibfnamefont{A.}~\bibnamefont{Sboychakov}},
  \bibinfo{author}{\bibfnamefont{A.}~\bibnamefont{Rakhmanov}},
  \bibnamefont{and} \bibinfo{author}{\bibfnamefont{F.}~\bibnamefont{Nori}},
  {``}\bibinfo{title}{Electronic properties of graphene-based bilayer
  systems},{''} \bibinfo{journal}{Phys. Rep.} \textbf{\bibinfo{volume}{648}},
  \bibinfo{pages}{1 } (\bibinfo{year}{2016}).

\bibitem[{\citenamefont{Rice}(1970)}]{Rice}
\bibinfo{author}{\bibfnamefont{T.M.} \bibnamefont{Rice}},
  {``}\bibinfo{title}{Band-Structure Effects in Itinerant
  Antiferromagnetism},{''} \bibinfo{journal}{Phys. Rev. B}
  \textbf{\bibinfo{volume}{2}}, \bibinfo{pages}{3619} (\bibinfo{year}{1970}).

\bibitem[{\citenamefont{Gorbatsevich et~al.}(1992)\citenamefont{Gorbatsevich,
  Kopaev, and Tokatly}}]{tokatly1992}
\bibinfo{author}{\bibfnamefont{A.}~\bibnamefont{Gorbatsevich}},
  \bibinfo{author}{\bibfnamefont{Y.}~\bibnamefont{Kopaev}}, \bibnamefont{and}
  \bibinfo{author}{\bibfnamefont{I.}~\bibnamefont{Tokatly}},
  {``}\bibinfo{title}{Band theory of phase stratification},{''}
  \bibinfo{journal}{Zh. Eksp. Teor. Fiz.} \textbf{\bibinfo{volume}{101}},
  \bibinfo{pages}{971} (\bibinfo{year}{1992}), \bibinfo{note}{[Sov. Phys. JETP
  {\bf 74}, 521 (1992)]}.

\bibitem[{\citenamefont{Eremin and Chubukov}(2010)}]{eremin_chub2010}
\bibinfo{author}{\bibfnamefont{I.}~\bibnamefont{Eremin}} \bibnamefont{and}
  \bibinfo{author}{\bibfnamefont{A.V.} \bibnamefont{Chubukov}},
  {``}\bibinfo{title}{Magnetic degeneracy and hidden metallicity of the
  spin-density-wave state in ferropnictides},{''} \bibinfo{journal}{Phys. Rev.
  B} \textbf{\bibinfo{volume}{81}}, \bibinfo{pages}{024511}
  (\bibinfo{year}{2010}).

\bibitem[{\citenamefont{Rakhmanov et~al.}(2012)\citenamefont{Rakhmanov,
  Rozhkov, Sboychakov, and Nori}}]{PrlOur}
\bibinfo{author}{\bibfnamefont{A.L.}~\bibnamefont{Rakhmanov}},
  \bibinfo{author}{\bibfnamefont{A.V.} \bibnamefont{Rozhkov}},
  \bibinfo{author}{\bibfnamefont{A.O.} \bibnamefont{Sboychakov}},
  \bibnamefont{and} \bibinfo{author}{\bibfnamefont{F.}~\bibnamefont{Nori}},
  {``}\bibinfo{title}{Instabilities of the $AA$-Stacked Graphene Bilayer},{''}
  \bibinfo{journal}{Phys. Rev. Lett.} \textbf{\bibinfo{volume}{109}},
  \bibinfo{pages}{206801} (\bibinfo{year}{2012}).

\bibitem[{\citenamefont{Sboychakov
  et~al.}(2013{\natexlab{a}})\citenamefont{Sboychakov, Rozhkov, Rakhmanov, and
  Nori}}]{PrbOur}
\bibinfo{author}{\bibfnamefont{A.O.}~\bibnamefont{Sboychakov}},
  \bibinfo{author}{\bibfnamefont{A.V.}~\bibnamefont{Rozhkov}},
  \bibinfo{author}{\bibfnamefont{A.L.}~\bibnamefont{Rakhmanov}},
  \bibnamefont{and} \bibinfo{author}{\bibfnamefont{F.}~\bibnamefont{Nori}},
  {``}\bibinfo{title}{Antiferromagnetic states and phase separation in doped
  AA-stacked graphene bilayers},{''} \bibinfo{journal}{Phys. Rev. B}
  \textbf{\bibinfo{volume}{88}}, \bibinfo{pages}{045409}
  (\bibinfo{year}{2013}{\natexlab{a}}).

\bibitem[{\citenamefont{Sboychakov
  et~al.}(2013{\natexlab{b}})\citenamefont{Sboychakov, Rakhmanov, Rozhkov, and
  Nori}}]{PrbROur}
\bibinfo{author}{\bibfnamefont{A.O.} \bibnamefont{Sboychakov}},
  \bibinfo{author}{\bibfnamefont{A.L.}~\bibnamefont{Rakhmanov}},
  \bibinfo{author}{\bibfnamefont{A.V.}~\bibnamefont{Rozhkov}},
  \bibnamefont{and} \bibinfo{author}{\bibfnamefont{F.}~\bibnamefont{Nori}},
  {``}\bibinfo{title}{Metal-insulator transition and phase separation in doped
  AA-stacked graphene bilayer},{''} \bibinfo{journal}{Phys. Rev. B}
  \textbf{\bibinfo{volume}{87}}, \bibinfo{pages}{121401}
  (\bibinfo{year}{2013}{\natexlab{b}}).

\bibitem[{\citenamefont{Sboychakov
  et~al.}(2013{\natexlab{c}})\citenamefont{Sboychakov, Rozhkov, Kugel,
  Rakhmanov, and Nori}}]{Sboychakov_PRB2013_PS_pnict}
\bibinfo{author}{\bibfnamefont{A.O.} \bibnamefont{Sboychakov}},
  \bibinfo{author}{\bibfnamefont{A.V.}~\bibnamefont{Rozhkov}},
  \bibinfo{author}{\bibfnamefont{K.I.}~\bibnamefont{Kugel}},
  \bibinfo{author}{\bibfnamefont{A.L.}~\bibnamefont{Rakhmanov}},
  \bibnamefont{and} \bibinfo{author}{\bibfnamefont{F.}~\bibnamefont{Nori}},
  {``}\bibinfo{title}{Electronic phase separation in iron pnictides},{''}
  \bibinfo{journal}{Phys. Rev. B} \textbf{\bibinfo{volume}{88}},
  \bibinfo{pages}{195142} (\bibinfo{year}{2013}{\natexlab{c}}).

\bibitem[{\citenamefont{Akzyanov et~al.}(2014)\citenamefont{Akzyanov,
  Sboychakov, Rozhkov, Rakhmanov, and Nori}}]{PrbVOur}
\bibinfo{author}{\bibfnamefont{R.S.} \bibnamefont{Akzyanov}},
  \bibinfo{author}{\bibfnamefont{A.O.}~\bibnamefont{Sboychakov}},
  \bibinfo{author}{\bibfnamefont{A.V.}~\bibnamefont{Rozhkov}},
  \bibinfo{author}{\bibfnamefont{A.L.}~\bibnamefont{Rakhmanov}},
  \bibnamefont{and} \bibinfo{author}{\bibfnamefont{F.}~\bibnamefont{Nori}},
  {``}\bibinfo{title}{$AA$-stacked bilayer graphene in an applied electric
  field: Tunable antiferromagnetism and coexisting exciton order
  parameter},{''} \bibinfo{journal}{Phys. Rev. B}
  \textbf{\bibinfo{volume}{90}}, \bibinfo{pages}{155415}
  (\bibinfo{year}{2014}).

\bibitem[{\citenamefont{Sboychakov et~al.}(2017)\citenamefont{Sboychakov,
  Rakhmanov, Kugel, Rozhkov, and Nori}}]{prb_sl2017}
\bibinfo{author}{\bibfnamefont{A.O.} \bibnamefont{Sboychakov}},
  \bibinfo{author}{\bibfnamefont{A.L.}~\bibnamefont{Rakhmanov}},
  \bibinfo{author}{\bibfnamefont{K.I.}~\bibnamefont{Kugel}},
  \bibinfo{author}{\bibfnamefont{A.V.}~\bibnamefont{Rozhkov}},
  \bibnamefont{and} \bibinfo{author}{\bibfnamefont{F.}~\bibnamefont{Nori}},
  {``}\bibinfo{title}{Magnetic field effects in electron systems with imperfect
  nesting},{''} \bibinfo{journal}{Phys. Rev. B} \textbf{\bibinfo{volume}{95}},
  \bibinfo{pages}{014203} (\bibinfo{year}{2017}).

\bibitem[{\citenamefont{Rakhmanov et~al.}(2013)\citenamefont{Rakhmanov,
  Rozhkov, Sboychakov, and Nori}}]{our_chrom2013}
\bibinfo{author}{\bibfnamefont{A.L.} \bibnamefont{Rakhmanov}},
  \bibinfo{author}{\bibfnamefont{A.V.}~\bibnamefont{Rozhkov}},
  \bibinfo{author}{\bibfnamefont{A.O.}~\bibnamefont{Sboychakov}},
  \bibnamefont{and} \bibinfo{author}{\bibfnamefont{F.}~\bibnamefont{Nori}},
  {``}\bibinfo{title}{Phase separation of antiferromagnetic ground states in
  systems with imperfect nesting},{''} \bibinfo{journal}{Phys. Rev. B}
  \textbf{\bibinfo{volume}{87}}, \bibinfo{pages}{075128}
  (\bibinfo{year}{2013}).

\bibitem[{\citenamefont{Igoshev et~al.}(2010)\citenamefont{Igoshev, Timirgazin,
  Katanin, Arzhnikov, and Irkhin}}]{IrkhinPRB2010}
\bibinfo{author}{\bibfnamefont{P.A.} \bibnamefont{Igoshev}},
  \bibinfo{author}{\bibfnamefont{M.A.}~\bibnamefont{Timirgazin}},
  \bibinfo{author}{\bibfnamefont{A.A.}~\bibnamefont{Katanin}},
  \bibinfo{author}{\bibfnamefont{A.K.}~\bibnamefont{Arzhnikov}},
  \bibnamefont{and} \bibinfo{author}{\bibfnamefont{V.Y.}
  \bibnamefont{Irkhin}}, {``}\bibinfo{title}{Incommensurate magnetic order and
  phase separation in the two-dimensional Hubbard model with nearest- and
  next-nearest-neighbor hopping},{''} \bibinfo{journal}{Phys. Rev. B}
  \textbf{\bibinfo{volume}{81}}, \bibinfo{pages}{094407}
  (\bibinfo{year}{2010}).

\bibitem[{\citenamefont{Lorenzana et~al.}(2001)\citenamefont{Lorenzana,
  Castellani, and Castro}}]{di_castro1}
\bibinfo{author}{\bibfnamefont{J.}~\bibnamefont{Lorenzana}},
  \bibinfo{author}{\bibfnamefont{C.}~\bibnamefont{Castellani}},
  \bibnamefont{and} \bibinfo{author}{\bibfnamefont{C.}
  \bibnamefont{Di~Castro}}, {``}\bibinfo{title}{Phase separation frustrated by the
  long-range Coulomb interaction. I. Theory},{''} \bibinfo{journal}{Phys. Rev.
  B} \textbf{\bibinfo{volume}{64}}, \bibinfo{pages}{235127}
  (\bibinfo{year}{2001}).

\bibitem[{\citenamefont{Bianconi et~al.}(2015)\citenamefont{Bianconi, Poccia,
  Sboychakov, Rakhmanov, and Kugel}}]{bianconi2015intrinsic}
\bibinfo{author}{\bibfnamefont{A.}~\bibnamefont{Bianconi}},
  \bibinfo{author}{\bibfnamefont{N.}~\bibnamefont{Poccia}},
  \bibinfo{author}{\bibfnamefont{A.}~\bibnamefont{Sboychakov}},
  \bibinfo{author}{\bibfnamefont{A.}~\bibnamefont{Rakhmanov}},
  \bibnamefont{and} \bibinfo{author}{\bibfnamefont{K.}~\bibnamefont{Kugel}},
  {``}\bibinfo{title}{Intrinsic arrested nanoscale phase separation near a
  topological Lifshitz transition in strongly correlated two-band metals},{''}
  \bibinfo{journal}{Superconductor Science and Technology}
  \textbf{\bibinfo{volume}{28}}, \bibinfo{pages}{024005}
  (\bibinfo{year}{2015}).

\end{thebibliography}

\end{document}